\documentclass[12pt]{article}
\usepackage{graphicx}
\usepackage{mathptmx,mathrsfs}
\usepackage{amsfonts,amsmath,amssymb,amsthm}
\usepackage{indentfirst}
\usepackage{cite}
\usepackage{hyperref}
\usepackage{enumitem}
\usepackage{bm,graphics,color}

\numberwithin{equation}{section}
\theoremstyle{definition}

\begin{document}

\title{Higgs Mechanism and Debye Screening in the Generalized Electrodynamics}

\author{C. A. Bonin$^{1}$\thanks{carlosbonin@gmail.com}, G. B. de Gracia$^{2}$\thanks{gb9950@gmail.com}, A. A. Nogueira$^{3}$\thanks{andsogueira@hotmail.com (corresponding author)},\, and B. M. Pimentel$^{2}$\thanks{b.m.pimentel@gmail.com}
\\
\textit{$^{1}$}\textit{\small{}Department of Mathematics \& Statistics, State University of Ponta Grossa (DEMAT/UEPG),}\\ \textit{\small{} Avenida Carlos Cavalcanti 4748,84030-900 Ponta Grossa, PR, Brazil}\\
\textit{$^{2}$}\textit{\small{} Instituto de F\' isica Te\' orica (IFT), Universidade Estadual Paulista}\\
\textit{\small{} Rua Dr. Bento Teobaldo Ferraz 271, Bloco II Barra Funda, CEP
01140-070 S\~ao Paulo, SP, Brazil}\\
\textit{$^{3}$}\textit{\small{}Universidade Federal de Goi\' as, Instituto de F\'isica, Av. Esperan\c{c}a, 74690-900, Goi\^ania, Goi\' as, Brasil}}
\maketitle
\date{}

\begin{abstract}
In this work we study the Higgs mechanism and the Debye shielding for the Bopp-Podolsky theory of electrodynamics. We find that not only the massless sector of the Podolsky theory acquires a mass in both these phenomena, but also that its massive sector has its mass changed. Furthermore, we find a mathematical analogy in the way these masses change between these two mechanisms. Besides exploring the behaviour of the screened potentials, we find a temperature for which the presence of the generalized gauge field may be experimentally detected.
\end{abstract}
\newpage{}

\section{Introduction}

The mechanisms behind mass generation for particles play a crucial role in the advancement of our understanding of the fundamental forces of Nature. For instance, our current description of three out of five fundamental interactions is through the action of gauge fields \cite{Rai}. This idea has one of its origins in Nuclear Physics with the Yukawa theory  wherein the screened Coulomb potential arises from the exchange of a massive pion between nucleons \cite{Yukawa}. Since the field mediator in the Yukawa theory has a nonvanishing mass, the corresponding force has a finite range. Actually, the amplitude of the classical Yukawa potential in the static limit decays exponentially with the distance when compared with the Coulomb potential and the range of this interaction, \textit{id est}, its typical length, is inversely proportional to the mass of the mediator particle.

A not altogether dissimilar idea arose independently in a completely different physical context: the description of the Meissner-Oschenfeld effect in low-temperature superconductivity \cite{Meissner}. In this phenomenon, weak magnetic fields are expelled from superconductors when their temperatures are bellow their respective critical temperatures. One of the first attempts for a theoretical description for this effect was given by the London Brothers' theory \cite{London}. The London Brothers hypothesised  the electric current $\mathbf{j}$ in a superconductor bellow its critical temperature is proportional to the gauge field $\mathbf{A}$. This assumption leads to a Helmholtz equation for each component of the static magnetic field inside the superconductor whose solutions comprise an exponential decay with a typical length - called London penetration length - depending on the characteristics of the particular material. So, in the London theory, magnetic fields do penetrate the superconductors, but they are damped  exponentially with the distance. As a matter of fact, the London Brothers' general results can also be achieved by assuming that, inside a superconductor bellow its critical temperature, the electromagnetic field acquires an effective mass inversely proportional to the London penetration length. In other words, the electromagnetic field inside a superconductor is more easily described not in terms of the fundamental Maxwell theory but in terms of an effective Proca field whose mass depends on the particular characteristics of the superconductor \cite{Proca}. A refined theory for superconductivity was proposed by Ginzburg and Landau and it is called the phenomenological Ginzburg-Landau theory \cite{Ginzburg Landau}. The Ginzburg-Landau theory consists of a nonrelativistic self-interacting charged scalar field interacting with a magnetic field. In modern terminology, their phenomenological theory can be described as follows. Due to the self-interacting potential of the scalar field the theory is gauge-invariant above the superconductor's critical temperature and, as usual, the electromagnetic field is massless. However, a phase transition takes place at the critical temperature in such a way that bellow that temperature the potential effectively breaks the gauge symmetry, giving rising to a non zero mass for the electromagnetic field. This, in turn, makes the effective description of the electromagnetic field be made through a Proca field, giving similar results to those of the London Brothers' theory for the explanation of the Meissner-Oschenfeld effect. This effective breaking of a symmetry due to a scalar potential is essentially the Anderson-Englert-Brout-Higgs-Guralnik-Hagen-Kibble mechanism.

The Anderson-Englert-Brout-Higgs-Guralnik-Hagen-Kibble mechanism (or Higgs mechanism, for short) for symmetry breaking was first proposed by Anderson in the nonrelativistic context based on plasma physics and similarities with superconductivity theories and it was later extended simultaneously by Englert and Brout, Higgs, and Guralnik, Hagen, and Kibble for relativistic fields \cite{Anderson, Englert, Higgs1, Guralnik, Rub}. The mechanism corresponds to a self-interacting charged scalar field interacting with a gauge field. Depending on the sign of a parameter, the self-interacting potential may have either a single minimum value or infinitely many minima. In the former case the theory is gauge-invariant and, as a consequence, the gauge field is massless. In the latter, a spontaneous symmetry breaking is said to occur and the gauge field acquires a mass. This mechanism plays a key role in the theory of the electroweak interaction. A prototype theory assuming that the weak coupling between fermions could be due to the exchange of massive vector bosons was first introduced by Lopes \cite{JLLopes}. After the advent
of the Higgs mechanism of symmetry breaking, the unification of weak and electromagnetic interactions was fulfilled by Glashow, Salam and Ward, and Weinberg which resulted in a massless vector field for the Abelian sector of the interaction and three massive vector gauge fields for the $SU\left(2\right)$ one \cite{Glashow, Salam Ward, Weinberg}. This furnished a more precise description of the electroweak interaction than the phenomenological theory by Fermi \cite{Fermi}. A new way of studying spontaneous symmetry breaking was obtained with the effective action technique introduced by Coleman and Weinberg \cite{Coleman}. The effective action technique opened up the possibility to explore the Jackiw-Dolan concept of symmetry restoration when the temperature of the system increases above a critical temperature \cite{JD}. This effect has its philosophical roots in the phenomenological Ginzburg-Landau theory of superconductivity.

Meanwhile, another revolution in the way we think of effective masses of gauge fields was taking place in Plasma Physics: the electric field shielding, also known as Debye screening. In order to grasp the essence of this phenomenon, let us picture a gas of classical electrons randomly but (in large scale) homogeneously  displaced spacewise under the assumption of charge neutrality, that is, that there exists a background of fixed ions for this gas in such a way that the total net electric charge is zero. If we insert a negative electric charge $q$ in this gas it will repel the electrons in its vicinity due to the Coulomb force. So, there is a region around the test charge where no electron (or very few of them) can be found. Since the gas is assumed to be overall homogeneous, this region with fewer electrons can be seen from afar as a region with a positive electric charge concentration due the ionic background. Measuring the field generated by the test charge away from it will show that it is shielded, or screened. This pictorial view can be made rigorous when considering a mathematical model for the system. This was first done by Debye and H\"{u}ckel \cite{Debye}. When they computed the effective Coulomb potential away from the test charge, they found that it is exponentially damped quantitatively just like the Yukawa potential and qualitatively like the London magnetic fields discussed above. The typical length of this screened potential is known as Debye length and its inverse, as in the previous cases, is proportional to a mass for the gauge field. This is known as Debye mass. The importance of this effect cannot be understated, as it is one of the simplest examples of renormalization. The Debye screening can be viewed as a process of renormalization for the test charge in the Coulomb potential as $q\rightarrow q e^{-m_D r}$ (in natural units) where $m_D$ is the Debye mass and $r$ the distance to the test charge \cite{McComb}. The Debye screening has been a topic of great implications in plasma physics (either classical or quantum, and for both nonrelativistic and relativistic), electrolytes and colloids \cite{Fetter Walecka, LeBe, Yang, Landau}.

Unlike the case of the Higgs mechanism, the presence of the Debye mass in a gauge field does not necessarily break the gauge symmetry of the theory.\footnote{This can be verified by noticing that the Ward-Fradkin-Takahashi identities are still satisfied in thermodynamic equilibrium for, say, quantum electrodynamics \cite{Fradkin}} We have plenty of examples where massive gauge fields, contrary of popular belief, maintain gauge symmetries intact. The first of these examples was studied by Schwinger where a dynamical mass generation for the quantum gauge field does take place, but the transversality of polarization tensor shows that theory is still gauge invariant \cite{Schwinger}. Another example of presence of mass in a gauge field which exhibits gauge symmetry can be found in theories of higher-order derivatives, from which the theory proposed by Bopp and Podolsky is the prime example \cite{Cuzinatto,Bop}. The Bopp-Podolsky theory for the electromagnetic field, also known as generalized electromagnetism - or generalized electrodynamics in the case where the dynamics of the sources of the gauge field are taking into consideration - is the main theory we study in this paper.  The Lagrangian density for this theory has a term involving second-order derivatives of the gauge field and possesses a free parameter that can be identified with a mass, called Podolsky mass $m_P$. This generalized electrodynamics gives the correct finite expressions for self-force of charged particles, as shown by Frenkel and Zayats \cite{Frank}. Generally speaking, Podolsky electrodynamics presents a better ultraviolet behavior than Maxwell's, in a way closely related to the Pauli-Villars-Rayski regularization scheme \cite{PV}. Despite the origin of the Podolsky mass remaining a mistery, several limits for its value have been set on experimental grounds through the years, with better and better accuracy for its lower bound \cite{Bonin, Bufalorenormalizado, Daniel}. Podolsky electrodynamics can reproduce most of the results of Maxwell theory, but it is worth noticing that it breaks the dual symmetry \cite{Brandt}. The generalized electromagnetic field has five degrees of freedom, two of them associated with the usual Maxwell field while the other three comes from a Proca field \cite{And}. This result has been corroborated by statiscal analysis \cite{Bonin}. In the free case, Podolsky theory is equivalent to Lee-Wick theory wherein the gauge field is complex \cite{LeeW}.

In the present paper we study the implications of the Higgs mechanism for electrostatic fields and the Debye screening in the context of the Podolsky electrodynamics. In section \ref{Classical Podosky Electrodynamics} we review the basics of the generalized theory with focus on its electrostatic potential. In section \ref{Higgs Mechanism} we consider the essence of the Higgs mechanism for the Bopp-Podolsky electromagnetism and study its implications for the theory's electrostatic field. In section \ref{Debye Screening} we consider the Podolsky plasma and compute (an approximation for) the Debye mass. In section \ref{beyond} we consider a region of parameters where new drastically Physics is expected to be found. In section \ref{Conclusions} we present our final thoughts on this subject. Throughout this paper we use the natural unit systems in which $\hbar=1$, $c=1$, $k_B=1$, and the Minkowski metric signature used is $\left(+,-,-,-\right)$.


\section{The Electrostatic Potential in the Generalized Classical Electrodynamics}\label{Classical Podosky Electrodynamics}

In this section, we review some aspects of the Classical Podolsky
Electrodynamics. We start by writing down the Lagrangian density for the theory in
$(3+1)$ dimensions as
\begin{equation}
\mathcal{L}_{CP}  =  -\frac{1}{4}\mathcal{F}_{\mu\nu}\mathcal{F}^{\mu\nu} +\frac{1}{2m_{P}^{2}}\partial^{\mu}\mathcal{F}_{\mu\nu}\partial_{\xi}\mathcal{F}^{\xi\nu} -\mathcal{J}^{\mu}\mathcal{A}_{\mu},\label{first Lagrangian}
\end{equation}
where $\mathcal{F}_{\mu\nu}  \equiv  \partial_{\mu}\mathcal{A}_{\nu}-\partial_{\nu}\mathcal{A}_{\mu}$ is the field-strength, $\mathcal{A}$ is the electromagnetic (or Podolsky) field, $m_{P}>0$ is the Podolsky mass, and $\mathcal{J}$ is an external source for the gauge field.

From (\ref{first Lagrangian}) we get the following Euler-Lagrange equations:
\begin{eqnarray}
\left(\frac{\square}{m_{P}^{2}}+1\right)\partial_{\mu}\mathcal{F}^{\mu\nu} & = & \mathcal{J}^{\nu},
\end{eqnarray}
where $\square\equiv\partial_{\mu}\partial^{\mu}$.

After imposing the generalized Lorenz condition \cite{Galvao}
\begin{eqnarray}
\left(\frac{\square}{m_{P}^{2}}+1\right)\partial_{\mu}\mathcal{A}^{\mu} & = & 0
\end{eqnarray}
the equations of motion become simplified
\begin{eqnarray}
\left(\frac{\square}{m_{P}^{2}}+1\right)\square\mathcal{A}^{\mu} & = & \mathcal{J}^{\mu}.\label{equations of motion}
\end{eqnarray}

For our purposes, we are interested in the static regime:
\begin{eqnarray}
\left(\frac{\nabla^{2}}{m_{P}^{2}}-1\right)\nabla^{2}\mathcal{A}^{\mu} & = & \mathcal{J}^{\mu}.
\end{eqnarray}

The solution for this equation is
\begin{eqnarray}
\mathcal{A}^{\mu}\left(\mathbf{x}\right) & = & \intop_{V}d^{3}y\, G_{P}\left(\mathbf{x}-\mathbf{y}\right)\mathcal{J}^{\mu}\left(\mathbf{y}\right),
\end{eqnarray}
where $V$ is any volume that encompasses $\mathcal{J}$ and $G_{P}$
is the Green function for the operator $\left(\frac{\nabla^{2}}{m_{P}^{2}}-1\right)\nabla^{2}$.

In particular, for a point electric charge lying in the origin of
the coordinate system,
\begin{eqnarray}
\mathcal{J}^{0}\left(\mathbf{y}\right) & = & Q\delta^{\left(3\right)}\left(\mathbf{y}\right),\label{point charge}
\end{eqnarray}
where $Q\neq0$ is the electric charge value, the electrostatic potential
is
\begin{equation}
\mathcal{A}^{0}\left(\mathbf{x}\right)= \frac{Q}{4\pi\left|\mathbf{x}\right|}\left(1-e^{-m_{P}\left|\mathbf{x}\right|}\right).\label{electrostatic potential in Podolsky}
\end{equation}

The Podolsky electrostatic potential has some known features like the fact that it goes to the Maxwell's result in the limit $m_P\rightarrow + \infty$ and it presents a finite limit at short distances:

\begin{equation}\label{finite limit}
  \lim_{|\mathbf{x}|\rightarrow 0^+}\mathcal{A}^{0}\left(\mathbf{x}\right)=\frac{Q\, m_P}{4\pi}.
\end{equation}

In order to appreciate fully the physical content of the electrostatic potential (\ref{electrostatic potential in Podolsky}), let us consider again equations (\ref{equations of motion}), this time in the free case $\mathcal{J}\equiv 0$. In that situation, the equations of motion can be written as

\begin{equation}
 \left(\square+m_{P}^{2}\right)\square\mathcal{A}^{\mu}=0,
\end{equation}
whose solution is $\mathcal{A}^{\mu}=\mathcal{A}_M^{\mu}-\mathcal{A}_P^{\mu}$. Here, $\mathcal{A}_M^{\mu}$ is a massless vector field while $\mathcal{A}_P^{\mu}$ is a Proca field with mass $m_P$. This is connected with the five degrees of freedom of the generalized theory mentioned earlier. $\mathcal{A}^\mu_M$ and $\mathcal{A}_P^\mu$ are called the massless and the massive \textit{sectors} of the theory, respectively. Furthermore, (\ref{electrostatic potential in Podolsky}) can now be understood as a Yukawa potential subtracted from the usual Coulomb potential. So, a part of Podolsky electrostatic field is naturally shielded due to the Podolsky length $m_P^{-1}$, but the interaction is still long-ranged since it goes asymptotically back to the Maxwell's result for $|\mathbf{x}|\gg m_P^{-1}$.

Our next move will be to check how this electrostatic potential is
affected by the presence of an extra mass in the Podolsky field due
to the spontaneous symmetry breaking of the gauge symmetry.


\section{The Classical Podolsky Field and the Higgs Mechanism}\label{Higgs Mechanism}

In this section we shall see how the spontaneous symmetry breaking induced by the Higgs potential in classical field theory affects the Podolsky field. The mass generation for the gauge field due to the Higgs mechanism can be studied in the Podolsky theory \textit{via} the following Lagrangian density
\begin{eqnarray}
\mathcal{L}_{PH} & = & \mathcal{L}_{P}+\mathcal{L}_{H},\label{L PH-1}
\end{eqnarray}
where $\mathcal{L}_{P}$ is the Lagrangian density (\ref{first Lagrangian}) with $\mathcal{J}^\mu=0$ and
\begin{eqnarray}
\mathcal{L}_{H} & \equiv & \left(D_{\mu}\phi\right)^{*}D^{\mu}\phi-U\left(\left| \phi\right| \right);\\
D_{\mu} & \equiv & \partial_{\mu}+iq_{s}\mathcal{A}_{\mu};\\
U\left(\left| \phi\right| \right) & \equiv & -\kappa\left| \phi\right| ^{2}+\lambda\left| \phi\right|^{4}.
\end{eqnarray}

Here, $\phi$ is a complex  Poincaré scalar field  (and $\phi^{*}$ its
conjugate) called the Higgs field, $\kappa$, $\lambda$, and $q_{s}$ are real parameters with $\lambda>0$ and $q_{s}$
being the electric charge of the scalar field. $\mathcal{L}_{PH}$ is
$U\left(1\right)$-gauge invariant.

 For $\kappa\leq0$, the (unique) global minimum of the potential $U$ is $\phi=\phi^{*}=0$. This is the usual generalized classical   scalar electrodynamics: a self-interacting complex scalar field interacting also with the ordinary Podolsky gauge field. On the other hand, if $\kappa>0$, $\left|\phi\right|=0$ becomes a local maximum of the potential (which means $\phi=0$ is no longer a stable field configuration), while there are now infinitely, uncountably many minima, all of them satisfying
\begin{eqnarray}
\left| \phi\left(x\right)\right| & = & \sqrt{\frac{\kappa}{2\lambda}}.
\end{eqnarray}

This is the situation we are interested in in the present paper. Due to the global $U\left(1\right)$ symmetry of (\ref{L PH-1}), any field such that $\phi\left(x\right)=e^{i\alpha}\sqrt{\kappa/2\lambda}$, with $\alpha\in\mathbb{R}$, minimizes $U$. Our present goal is to expand (\ref{L PH-1}) around one of such minima. Due to the symmetry of the theory, it really does not matter which minimum we choose. So, without loss of generality, we shall expand the Lagrangian density around the value $\phi=\sqrt{\kappa/2\lambda}$. In order to do so, we rewrite the scalar field as

\begin{eqnarray}
\phi\left(x\right) & = & \frac{1}{\sqrt{2}}\left[\chi\left(x\right)+\sqrt{\frac{\kappa}{\lambda}}\right] e^{i\sqrt{\frac{\lambda}{\kappa}}\theta\left(x\right)}\label{new scalar fields}
\end{eqnarray}
and its complex conjugate accordingly. Here, $\chi$  and $\theta$ are real scalar
fields and $\inf_{x}\chi\left(x\right)=-\sqrt{\kappa/2\lambda}$. In order for us to understand the role of each of these new scalar fields, let us expand the term involving $\theta$ as

\begin{eqnarray}
e^{i\sqrt{\frac{\lambda}{\kappa}}\theta\left(x\right)} & = & 1+i\sqrt{\frac{\lambda}{\kappa}}\theta\left(x\right) +\mathcal{O}\left(\left(\sqrt{\frac{\lambda}{\kappa}}\right)^{2}\right).
\end{eqnarray}

Doing so approximates the potential $U$ to

\begin{eqnarray}
U\left(\left| \phi\right| \right)
  & \simeq & \kappa\left[\chi\left(x\right)\right]^{2}-\frac{\kappa^{2}}{4\lambda}+\lambda\left[\chi\left(x\right)\right]^{3}\sqrt{\frac{\kappa}{\lambda}} +\lambda\chi\left(x\right)\left[\theta\left(x\right)\right]^{3}\sqrt{\frac{\kappa}{\lambda}} +\frac{\lambda\left[\chi\left(x\right)\right]^{4}}{4}+\nonumber \\
 &  & +\frac{\lambda}{2}\left[\chi\left(x\right)\right]^{2}\left[\theta\left(x\right)\right]^{2}.
\end{eqnarray}

Since we are dealing with a positive parameter $\kappa$, the quadratic term in $\chi$ tells us that it is a massive field with mass $\sqrt{\kappa}$. Seeing that there is no correspondent term for $\theta$, the expression above shows us that $\theta$ is a massless scalar field. This is a consequence of the Goldstone theorem and, for that reason, $\theta$ is called the Goldstone boson of this theory.\footnote{The path we followed here is close to that of Goldstone's original paper \cite{Goldstone, JH}. A rigorous proof of the Goldstone theorem for classical fields (without gauge field, though) can be found in \cite{Strocchi}.}

Now, going back to the full representation (\ref{new scalar fields}), it is pretty clear that the potential $U$ is independent of the Goldstone boson. What is not so obvious is that the whole Lagrangian density (\ref{L PH-1}) does not depend on $\theta$ either. In order to prove that, we recall the Lagrangian density (\ref{L PH-1}) is invariant under local
gauge transformations and we perform the following gauge transformation:
\begin{eqnarray}
\mathcal{A}^{\mu}\left(x\right) & \rightarrow & \mathcal{A}^{\prime\mu}\left(x\right)=\mathcal{A}^{\mu}\left(x\right)+\partial^{\mu}\left[\frac{1}{q_{s}}\sqrt{\frac{\lambda}{\kappa}}\theta\left(x\right)\right];\\
\phi\left(x\right) & \rightarrow & \phi^{\prime}\left(x\right)=e^{iq_{s}\left[-\frac{1}{q_{s}}\sqrt{\frac{\lambda}{\kappa}}\theta\left(x\right)\right]}\phi\left(x\right);\\
\phi^{*}\left(x\right) & \rightarrow & \phi^{\prime*}\left(x\right)=e^{-iq_{s}\left[-\frac{1}{q_{s}}\sqrt{\frac{\lambda}{\kappa}}\theta\left(x\right)\right]}\phi^{*}\left(x\right).
\end{eqnarray}

These last two transformations are equivalent to
\begin{eqnarray}
\chi\left(x\right) & \rightarrow & \chi^{\prime}\left(x\right)=\chi\left(x\right);\\
\theta\left(x\right) & \rightarrow & \theta^{\prime}\left(x\right)=\theta\left(x\right)-\theta\left(x\right)=0\label{transformation of Goldstone}
\end{eqnarray}
in (\ref{new scalar fields}). By the very definition of covariant derivative, we have
\begin{eqnarray}
D^{\mu}\phi\left(x\right)
 & \rightarrow & e^{-i\sqrt{\frac{\lambda}{\kappa}}\theta\left(x\right)}D^{\mu}\phi\left(x\right)
\end{eqnarray}
which, thanks to the gauge transformation (\ref{transformation of Goldstone}), shows us that
\begin{eqnarray}
\left(D_{\mu}\phi\right)^{*}D^{\mu}\phi & \rightarrow &  D_{\mu}^{*}\frac{1}{\sqrt{2}}\left[\chi\left(x\right)+\sqrt{\frac{\kappa}{\lambda}}\right] D^{\mu}\frac{1}{\sqrt{2}}\left[\chi\left(x\right)+\sqrt{\frac{\kappa}{\lambda}}\right]\label{covariant transformation}
\end{eqnarray}
is independent of the Goldstone boson as well and so is the whole Lagrangian (\ref{L PH-1}), \textit{quod erat demonstrandum}. Furthermore, expliciting the covariant derivative yields
\begin{eqnarray}
\left(D_{\mu}\phi\right)^{*}D^{\mu}\phi & = & \frac{1}{2}\partial_{\mu}\chi\left(x\right)\partial^{\mu} \chi\left(x\right)+\frac{iq_{s}}{2}\partial_{\mu}\chi\left(x\right)\mathcal{A}^{\mu}\left(x\right)\chi\left(x\right)+
+\frac{iq_{s}}{2}\sqrt{\frac{\kappa}{\lambda}}\partial_{\mu}\chi\left(x\right) \mathcal{A}^{\mu}\left(x\right)+\nonumber\\
 &  & -\frac{iq_{s}}{2}\mathcal{A}_{\mu}\left(x\right)\chi\left(x\right) \partial^{\mu}\chi\left(x\right)-\frac{iq_{s}}{2}\sqrt{\frac{\kappa}{\lambda}} \mathcal{A}_{\mu}\left(x\right)\partial^{\mu}\chi\left(x\right)+\nonumber\\
 &  & +\frac{q_{s}^{2}}{2}\mathcal{A}_{\mu}\left(x\right) \chi\left(x\right)\mathcal{A}^{\mu}\left(x\right)\chi\left(x\right) +q_{s}^{2}\sqrt{\frac{\kappa}{\lambda}}\mathcal{A}_{\mu}\left(x\right)\chi\left(x\right) \mathcal{A}^{\mu}\left(x\right)+\nonumber\\
 &  & +\frac{m_H^2}{2}\mathcal{A}_{\mu}\left(x\right)\mathcal{A}^{\mu}\left(x\right).
\end{eqnarray}

Here, we have defined\footnote{Notice that $m_H$ is non negative.}

\begin{align}
  m_H &\,\equiv \sqrt{\frac{q_{s}^{2}\kappa}{\lambda}}.\label{mass from Higgs}
\end{align}

With this result, we can rewrite (\ref{L PH-1}) as\footnote{Of course, we can rewrite (\ref{L PH-1}) in the form (\ref{somando zero}) even for $\kappa\leq 0$. However, there are some problems. First of all, the mass (\ref{mass from Higgs}) becomes zero for $\kappa=0$ (which makes the whole point void) or purely imaginary for $\kappa<0$. This last issue may be more or less remedied by inverting the form (\ref{somando zero}) to $\left(\mathcal{L}_{P}-\frac{m_H^2}{2}\mathcal{A}^{\mu}\mathcal{A}_{\mu}\right) +\left(\mathcal{L}_{H}+\frac{m_H^2}{2}\mathcal{A}^{\mu}\mathcal{A}_{\mu}\right)$. Nevertheless, in that case, $\mathcal{L}_{H}+\frac{m_H^2}{2}\mathcal{A}^{\mu}\mathcal{A}_{\mu}$ would contain a term that depends only on the vector field, which defeats the purpose of writing it like this.}

\begin{eqnarray}
\mathcal{L}_{PH} & = & \left(\mathcal{L}_{P}+\frac{m_H^2}{2}\mathcal{A}^{\mu}\mathcal{A}_{\mu}\right) +\left(\mathcal{L}_{H}-\frac{m_H^2}{2}\mathcal{A}^{\mu}\mathcal{A}_{\mu}\right).\label{somando zero}
\end{eqnarray}

 $\mathcal{L}_{P}+\frac{m_H^2}{2}\mathcal{A}^{\mu}\mathcal{A}_{\mu}$ is a theory of a free vector field $\mathcal{A}$ (with second order derivatives) while $\mathcal{L}_{H}-\frac{m_H^2}{2}\mathcal{A}^{\mu}\mathcal{A}_{\mu}$ is a theory of a real, massive scalar field $\chi$ interacting with itself and with the vector field $\mathcal{A}$ without terms that depend only on the vector field. We are interested in the first of these two: the Lagrangian density concerning only the Podolsky field. In order to proceed, we shall add a source for this gauge field satisfying the continuity equation and write

\begin{eqnarray}
\mathcal{L}_{PH}^{\left(1\right)} & \equiv & -\frac{1}{4}\mathcal{F}_{\mu\nu}\mathcal{F}^{\mu\nu}+\frac{1}{2m_{P}^{2}}\partial^{\mu}\mathcal{F}_{\mu\nu}\partial_{\xi}\mathcal{F}^{\xi\nu}+\frac{m_{H}^{2}}{2}\mathcal{A}_{\mu}\mathcal{A}^{\mu}-\mathcal{J}^{\mu}\mathcal{A}_{\mu}.
\end{eqnarray}

The equations of motion which arrive from this Lagrangian density
are
\begin{eqnarray}
\left(\frac{\square}{m_{P}^{2}}+1\right)\partial_{\mu}\mathcal{F}^{\mu\nu}+m_{H}^{2}\mathcal{A}^{\nu} & = & \mathcal{J}^{\nu}.
\end{eqnarray}

The continuity of the external current implies $\partial_{\nu}\mathcal{A}^{\nu}=0$ which, in turn, simplifies the above Euler-Lagrange equations:

\begin{eqnarray}
\left[\left(\frac{\square}{m_{P}^{2}}+1\right)\square+m_{H}^{2}\right]\mathcal{A}^{\mu} & = & \mathcal{J}^{\mu}.
\end{eqnarray}

The static regime equations can be obtained from this last expression and they read
\begin{equation}
\left[\left(\frac{\nabla^2}{m_{P}^{2}}-1\right)\nabla^2+m_{H}^{2}\right]\mathcal{A}^{\mu}  =  \mathcal{J}^{\mu},
\end{equation}
whose solution is

\begin{equation}
\mathcal{A}^{\mu}\left(\mathbf{x}\right)  =  \intop_{V}d^{3}y\, G_{PH}\left(\mathbf{x}-\mathbf{y}\right)\mathcal{J}^{\mu}\left(\mathbf{y}\right).
\end{equation}

In this equation, $G_{PH}$ is the Green function for the differential operator $\left(\frac{\nabla^2}{m_{P}^{2}}-1\right)\nabla^2+m_{H}^{2}$. In order to find the new electrostatic potential, we consider a point charge like (\ref{point charge}) with all other components vanishing. The resulting potential is

\begin{equation}
\mathcal{A}^{0}\left(\mathbf{x}\right) = \frac{Qm_{P}^{2}}{4\pi\left|\mathbf{\mathbf{x}}\right|\left(m_{H+}^{2}-m_{H-}^{2}\right)} \left(e^{-m_{H-}\left|\mathbf{\mathbf{x}}\right|} -e^{-m_{H+}\left|\mathbf{\mathbf{x}}\right|}\right),\label{electrostatic Higgs}
\end{equation}
where we have defined the masses\footnote{The results presented in this section are valid for $m_H\leq m_P/2$. As a consequence, not only both masses $m_{H+}$ and $m_{H-}$ are real but they are also both non negative.}

\begin{equation}
m_{H\pm}  \equiv  \sqrt{\frac{m_{P}^{2}\pm m_{P}\sqrt{m_{P}^{2}-4m_{H}^{2}}}{2}}.\label{mH plus minus}
\end{equation}

First of all, we see that if we turn the interaction between the Higgs field and the Podolsky field off, these masses go to

\begin{eqnarray}
  \lim_{m_H\rightarrow 0^+} m_{H+} &=& m_P;\label{limit mH going to zero mHplus} \\
  \lim_{m_H\rightarrow 0^+} m_{H-} &=& 0.\label{limit mH going to zero mHminus}
\end{eqnarray}

This shows that in this limit the electrostatic potential (\ref{electrostatic Higgs}) from the massive Podolsky field goes back to the usual Podolsky electrostatic potential (\ref{electrostatic potential in Podolsky}). Furthermore, the result (\ref{electrostatic Higgs}) and the limits above show that the massless sector of Podolsky acquires a mass $m_{H-}$ through the Higgs mechanism. This is not surprising, since the same phenomenon takes place when one is studying the Higgs field coupled with Maxwell's. What is really unexpected is that the Higgs mechanism alters the mass of the massive sector of the generalized gauge field from the Podolsky mass $m_P$ to a new value $m_{H+}<m_P$. In addition, it is worth noticing that the number of degrees of freedom of the theory (without sources) remains unchanged. In the Lagrangian density (\ref{L PH-1}) we had five degrees of freedom for the Podolsky field (being two associated with the massless sector of the theory and three with the massive one) and two for the complex scalar field. So, the initial degrees of freedom of the theory were seven. For vanishing sources, we ended up with one real scalar field (which has one degree of freedom), three degrees of freedom for the massive sector of the Podolsky field (the massive sector continues to be massive despite having its mass value changed), and also three for the former  massless sector which now has acquired a mass. So, the Podolsky field after the symmetry breaking adds six degrees of freedom to the final theory which, as a consequence, equates the original seven. This happens due to the Higgs mechanism, which made the Goldstone boson disappear from the final Lagrangian density.

Secondly, there are a couple of limits of interest in (\ref{electrostatic Higgs}). At short distances, the electrostatic potential for the Podolsky field in the broken symmetry regime is finite

\begin{equation}
  \lim_{\left|\mathbf{x}\right|\rightarrow 0^+}\mathcal{A}^{0}\left(\mathbf{x}\right) = \frac{Q\, m_P^2}{4\pi\left(m_{H+}+m_{H-}\right)},\label{finite limit modified}
\end{equation}
which is reminiscent of (\ref{finite limit}). In addition, in the limit of equal masses $m_{H-}=m_{H+}$, which happens when the mass acquired through the Higgs mechanism $m_H$ goes to  half the value of the Podolsky mass $m_P$, the result is a Yukawa potential with typical length $l_Y\equiv\sqrt{2}/m_P$ and the electric charge $Q$ renormalized to $Q\left|\mathbf{x}\right|/l_Y$:

\begin{equation}
  \lim_{m_H\rightarrow \frac{m_P^-}{2}}\mathcal{A}^{0}\left(\mathbf{x}\right) = \frac{\left(\frac{Q\left|\mathbf{x}\right|}{l_Y}\right)e^{-\frac{\left|\mathbf{x}\right|}{l_Y}}}{4\pi \left|\mathbf{x}\right|}.\label{equal renormalized}
\end{equation}

Alternatively, this can be seen as a pure evanescent wave generated by the electric charge $Q$ rescaled to the charge linear density ${Q}/{l_Y}$

\begin{equation}
  \lim_{m_H\rightarrow \frac{m_P^-}{2}}\mathcal{A}^{0}\left(\mathbf{x}\right) = \frac{\left(\frac{Q}{l_Y}\right)e^{-\frac{\left|\mathbf{x}\right|}{l_Y}}}{4\pi }\label{equal evanescent}
\end{equation}
in a manner that resembles the London Brothers' result for magnetic fields inside superconductors \cite{London}. Moreover, this is precisely the result we get by starting with the condition of equal masses:

\begin{equation}
 \left.\mathcal{A}^{0}\left(\mathbf{x}\right)\right|_{m_{H+}=m_{H-}=\frac{m_P}{\sqrt{2}}} = \frac{Q\,m_P}{4\sqrt{2}\pi}e^{-\frac{m_P}{\sqrt{2}}\left|\mathbf{x}\right|}.\label{equal continuous}
\end{equation}

Equations (\ref{equal evanescent}) and (\ref{equal continuous}) show that the Bopp-Podolsky electrostatic potential under the Higgs mechanism is a continuous function of the Higgs mass $m_H$ from the left of half the Podolsky mass, that is, $m_P/2$.

Lastly, we see that in the region of parameters $m_H\ll m_P$, the modified masses (\ref{mH plus minus}) behave as

\begin{eqnarray}
  m_{H+} &=& m_P\left(1+\mathcal{O}\left(\left(\frac{m_H}{m_P}\right)^2\right)\right);\label{approximation mHplus} \\
   m_{H-} &=& m_H\left(1+\mathcal{O}\left(\left(\frac{m_H}{m_P}\right)^2\right)\right).\label{approximation mHminus}
\end{eqnarray}

So, this is the region where our na\"{i}ve expectation is met with actualization: the massless sector of the gauge field acquires the mass $m_H$ obtained directly from the Higgs mechanism  and the massive sector retains its original mass $m_P$. In general, however, both masses are modified.

Next we shall study the Debye screening in the Generalized Quantum Plasma and compare the results with those shown in this section.


\section{Debye Screening in the Generalized Quantum Electrodynamics}\label{Debye Screening}

In this section we are interested in learning about the electrostatic field generated by an electric point charge not in the vacuum, but in a generalized relativistic quantum  plasma. A generalized relativistic quantum  plasma is nothing more than the Podolsky quantum electrodynamics in thermodynamic equilibrium \cite{BB}. Inside the plasma, under the assumptions of validity of the finite-temperature linear response theory, the  thermal expectation of the disturbed gauge field $\left\langle\delta\widehat{A}_\mu\left({x}\right)\right\rangle\equiv\mathcal{A}_\mu\left({x}\right)$ due to the presence of an external classical current density $\mathcal{J}_\nu\left(\mathbf{y}\right)$ is given by \cite{Fetter Walecka, LeBe,Yang}

\begin{equation}
\mathcal{A}_\mu\left(x\right)=\int d^4y\mathcal{D}^R_{\mu\nu}\left(x-y\right) \mathcal{J}^\nu\left(y\right). \label{potential for Debye}
\end{equation}

In this expression, $\mathcal{D}^R_{\mu\nu}\left(\cdot\right)$ is the retarded propagator of the (quantum) Podolsky gauge field in the plasma. Since we are interested in the electrostatic potential, we have $\mathcal{A}=\left(\mathcal{A}_0,\mathbf{0}\right)$ and $\mathcal{J}=\left(\mathcal{J}^0,\mathbf{0}\right)$, with $\mathcal{J}^0$ given by (\ref{point charge}), which imply the only component of the retarded propagator we need is that with $\mu=\nu=0$. The way to find the retarded propagator inside the medium is computing the electromagnetic Green function $\widetilde{\mathcal{D}}_{\mu\nu}\left(k^{Bn}\right)$ in thermodynamic equilibrium in the Fourier space and, then, Wick-rotating the variables as\footnote{We use the following notation for the theory in thermodynamic equilibrium: $k^{Bn}=\left(\omega_n^B,\mathbf{k}\right)$ and $k^{Fn}=\left(\omega_n^F,\mathbf{k}\right)$, where $\omega_n^B=2n\pi/\beta$ is the Bosonic Matsubara frequency and $\omega_n^F=\omega_n^B+\pi/\beta$ the Fermionic Matsubara frequency. In both cases, $\beta$ is the multiplicative inverse of the temperature and $n\in\mathbb{Z}$. Moreover, all the implicit summations are done with the Euclidean metric tensor $\delta_{\mu\nu}$.}

\begin{equation}
i\omega_n^B\rightarrow k_0+i\eta,\label{Wick rotation}
\end{equation}
where $\eta$ is any positive number and, in the end, computing the one-side limit $\eta\rightarrow 0^+$ \cite{Baym, Fetter Walecka, LeBe}.

In order to find $\widetilde{\mathcal{D}}_{\mu\nu}\left(k^{Bn}\right)$, we first need to compute a component of the polarization tensor, whose general form in thermodynamic equilibrium is\footnote{As a matter of fact, as shown in \cite{BB}, the most general form of the polarization tensor is $\widetilde{\Pi}_{\mu\nu}\left(k^{Bn}\right)  =  A\left(k^{Bn}\right)\left[\delta_{\mu\nu}-\frac{k_{\mu}^{Bn}k_{\nu}^{Bn}}{\left(k^{Bn}\right)^{2}}\right] +B\left(k^{Bn}\right)\left[\frac{k_{\mu}^{Bn}k_{\nu}^{Bn}}{\left(k^{Bn}\right)^{2}} -\left(\frac{k_{\mu}^{Bn}\delta_{\nu0}+k_{\nu}^{Bn}\delta_{\mu0}}{\omega_{n}^{B}}\right) +\frac{\left(k^{Bn}\right)^{2}\delta_{\mu0}\delta_{\nu0}}{\left(\omega_{n}^{B}\right)^{2}}\right] +I\left(k^{Bn}\right)\varepsilon_{0\mu\nu\xi}\frac{k_{\xi}^{Bn}}{\omega_{n}^{B}}$. Using (\ref{projector 1}), (\ref{projector 2}), and (\ref{projector 3}) we can show that we can rename the even scalar functions as $F\left(k^{Bn}\right)  \equiv  A\left(k^{Bn}\right)+\frac{\left|\mathbf{k}\right|^{2}}{\left(\omega_{n}^{B}\right)^{2}}B\left(k^{Bn}\right)$ and
$G\left(k^{Bn}\right)  \equiv  A\left(k^{Bn}\right)$. For our present analysis we can safely assume the odd function $I$ vanishes for two reasons: 1) the tensor $\varepsilon_{0\mu\nu\xi}\frac{k_{\xi}^{Bn}}{\omega_{n}^{B}}$ is not invertible. In order to see that, let us suppose, \textit{ad absurdum}, that a tensor $Q_{\mu\nu}\left(k^{Bn}\right)$ is its inverse. Then, by definition, we must have $Q_{\mu\rho}\left(k^{Bn}\right)\varepsilon_{0\rho\nu\xi}\frac{k_{\xi}^{Bn}}{\omega_{n}^{B}}=\delta_{\mu\nu}$. By choosing $\mu=\nu=0$ this expression teaches us that $0=1$, which is a contradiction. \textit{Ergo}, $\varepsilon_{0\mu\nu\xi}\frac{k_{\xi}^{Bn}}{\omega_{n}^{B}}$ is a singular tensor. 2) we only need to compute $\widetilde{\Pi}_{00}\left(k^{Bn}\right)$ anyway and there is no contribution of $I$ to it.}

\begin{eqnarray}
\widetilde{\Pi}_{\mu\nu}\left(k^{Bn}\right) & = & F\left(k^{Bn}\right)P_{\mu\nu}^{L}+G\left(k^{Bn}\right)P_{\mu\nu}^{T},\label{polarization}
\end{eqnarray}
where we have used the projectors \cite{LeBe}

\begin{eqnarray}
P_{00}^{T} & \equiv & P_{0i}^{T}=P_{i0}^{T}=0;\label{projector 1}\\
P_{ij}^{T} & \equiv & \delta_{ij}-\frac{k_{i}k_{j}}{\left|\mathbf{k}\right|^{2}};\label{projector 2}\\
P_{\mu\nu}^{L} & \equiv & \delta_{\mu\nu}-\frac{k_{\mu}^{Bn}k_{\nu}^{Bn}}{\left(k^{Bn}\right)^{2}}-P_{\mu\nu}^{T}.\label{projector 3}
\end{eqnarray}

From the polarization tensor (\ref{polarization}) we have the complete electromagnetic Green function in thermodynamic equilibrium (\textit{vide}  Fig. \ref{full}):

\begin{figure}
\centering
\includegraphics[width=10cm]{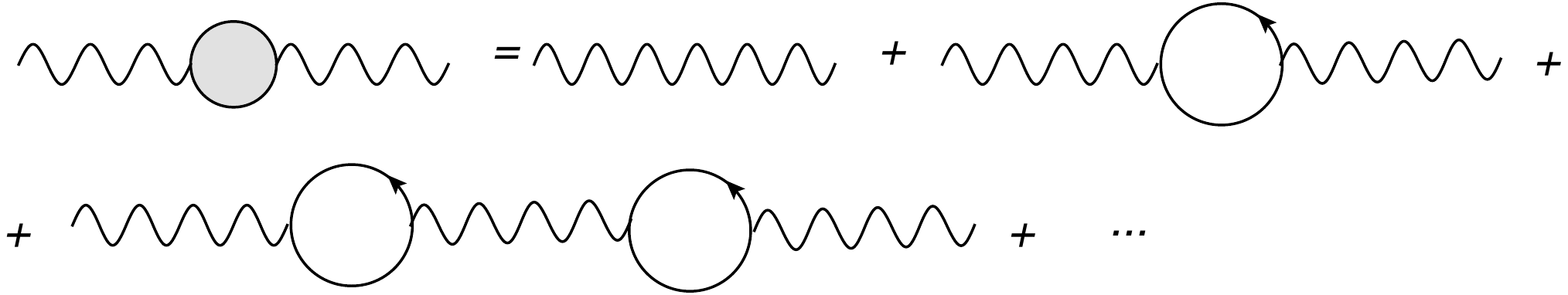}
\caption{Full thermal photon propagator}
\label{full}
\end{figure}

\begin{eqnarray}
\widetilde{\mathcal{D}}_{\mu\nu}\left(k^{Bn}\right)&=&\widetilde{D}_{\mu\nu}\left(k^{Bn}\right) +\widetilde{D}_{\mu\zeta}\left(k^{Bn}\right)\widetilde{\Pi}_{\zeta\kappa}\left(k^{Bn}\right)\widetilde{D}_{\kappa\nu}\left(k^{Bn}\right)\nonumber\\ &&+\widetilde{D}_{\mu\zeta}\left(k^{Bn}\right)\widetilde{\Pi}_{\zeta\kappa}\left(k^{Bn}\right) \widetilde{D}_{\kappa\lambda}\left(k^{Bn}\right)\widetilde{\Pi}_{\lambda\rho}\left(k^{Bn}\right)\widetilde{D}_{\rho\nu}\left(k^{Bn}\right)+...\cr\cr
&=& \frac{1}{\left\{ -\left[1+\frac{\left(k^{Bn}\right)^{2}}{m_{P}^{2}}\right]\left(k^{Bn}\right)^{2}+F\left(k^{Bn}\right)\right\} }P_{\mu\nu}^{L}+\frac{1}{\left\{ -\left[1+\frac{\left(k^{Bn}\right)^{2}}{m_{P}^{2}}\right]\left(k^{Bn}\right)^{2}+G\left(k^{Bn}\right)\right\} }P_{\mu\nu}^{T}+\cr\cr
&&+\frac{\alpha}{\left\{ -\left[1+ \frac{\left(k^{Bn}\right)^{2}}{m_{P}^{2}}\right]^2\left(k^{Bn}\right)^{2} \right\} }\frac{k_{\mu}^{Bn}k_{\nu}^{Bn}}{\left(k^{Bn}\right)^{2}}.\label{complete Green function}
\end{eqnarray}
in which the free generalized thermal electromagnetic propagator is written in the non-mixing gauge \cite{And}

\begin{equation}
\widetilde{D}_{\mu\nu}\left(k^{Bn}\right)=-\left[\frac{m_P^2}{\left(k^{Bn}\right)^2+m_P^2}\right] \left[\delta_{\mu\nu}+\left(\alpha-1\right)\frac{{k^{Bn}_{\mu}}{k^{Bn}_{\nu}}}{{\left(k^{Bn}\right)}^{2}}\right].
\end{equation}
For future use, it is worth noticing that

\begin{equation}
F\left(\mathbf{k},\omega_n^B\right)=\left[\frac{\left(\omega_n^B\right)^2+ \left|\mathbf{k}\right|^2}{\left|\mathbf{k}\right|^2}\right] \widetilde{\Pi}_{00}\left(\mathbf{k},\omega_n^B\right).\label{F and Pi}
\end{equation}

\begin{figure}
\centering
\includegraphics[width=6cm]{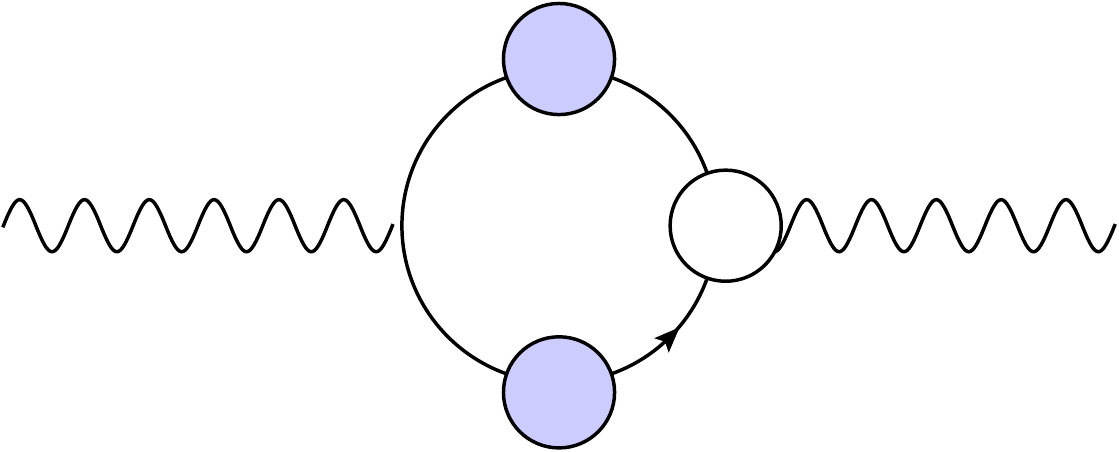}
\caption{Exact photon self-energy}
\label{full2}
\end{figure}

As it will become clear later, we need to compute $\lim_{\mathbf{k}\rightarrow \mathbf{0}}F\left(\omega_0^B,\mathbf{k}\right)$. According to the relation above, we need to find $\lim_{\mathbf{k}\rightarrow \mathbf{0}}\widetilde{\Pi}_{00}\left(0,\mathbf{k}\right)$. In order to compute this quantity, we use one of the Dyson-Schwinger-Fradkin equations (in the Fourier space, \textit{vide} Fig. \ref{full2})

\begin{eqnarray}
\widetilde{\Pi}_{\mu\nu}\left(\mathbf{k},\omega_{n}^{B}\right) & = & \sum_{l=-\infty}^{+\infty}\int\frac{d^{3}p}{\beta\left(2\pi\right)^{3}}q_{e}^{2}\left(\gamma_{\mu}^{E}\right)_{ab} \widetilde{\mathcal{S}}_{bc}\left(\mathbf{p},\omega_{l}^{F}\right)\widetilde{\Gamma}_{\nu\left(cd\right)} \left(\mathbf{p-k},\omega_{l-n}^{F};-\mathbf{k},\omega_{-n}^{B}\right)\nonumber\\
&&\times\widetilde{\mathcal{S}}_{da} \left(\mathbf{p-k},\omega_{l-n}^{F}\right).\label{Dyson Schwinger Fradkin}
\end{eqnarray}

In this expression, $q_e$ is the electron electric charge, $\left\{\gamma_{\mu}^{E}\right\}_{\mu\in\left\{0,1,2,3\right\}}$ are the Euclidean Dirac matrices, $\widetilde{\mathcal{S}}$ is the complete electron Green function, and $\widetilde{\Gamma}$ is the complete vertex function of the theory.

From the Ward identity in thermodynamic equilibrium\footnote{Notice that there is a typo with a missing $i$ in equation (165) of \cite{BB}.}

\begin{eqnarray}
ip_{\mu}^{Bl}\widetilde{\Gamma}_{\mu\left(ab\right)}\left(\mathbf{k},\omega_{n}^{F};\mathbf{p},\omega_{l}^{B}\right) & = & \widetilde{\mathcal{S}}_{ab}^{-1}\left(\mathbf{k}+\mathbf{p},\omega_{n+l}^{F}\right)-\widetilde{\mathcal{S}}_{ab}^{-1}\left(\mathbf{k},\omega_{n}^{F}\right),
\end{eqnarray}
we can deduce for $l=0$ and $\mathbf{p}\rightarrow \mathbf{0}$ \cite{Fradkin}
\begin{eqnarray}
\frac{\partial\widetilde{\mathcal{S}}_{ab}^{-1}\left(\mathbf{k},\omega_{n}^{F}\right)}{\partial\mu_{e}} & = & \widetilde{\Gamma}_{0\left(ab\right)}\left(\mathbf{k},\omega_{n}^{F};\mathbf{0},0\right).\label{Ward Fradkin}
\end{eqnarray}

Here, $\mu_e$ is the chemical potential associated with the conservation of the Noether charge

\begin{eqnarray}
\widehat{N} & = & \frac{1}{2}\left(\gamma_{0}^{E}\right)_{ab}\intop_{V}d^{3}z \left[\widehat{\overline{\psi}}_{a}\left(\mathbf{z}\right),\widehat{\psi}_{b}\left(\mathbf{z}\right)\right].\label{Noether}
\end{eqnarray}

From the definition of the Fermionic Green function and from equations (\ref{Dyson Schwinger Fradkin},\ref{Ward Fradkin},\ref{Noether}) as well we can write

\begin{eqnarray}
\lim_{\mathbf{k}\rightarrow\mathbf{0}}\widetilde{\Pi}_{00}\left(\mathbf{k},0\right)
 & = & -q_{e}^{2}\frac{\partial n\left(\beta, \mu_{e}\right)}{\partial\mu_{e}},\label{limit of Pi}
\end{eqnarray}
where

\begin{eqnarray}
n\left(\beta, \mu_{e}\right) & \equiv & \sum_{l=-\infty}^{+\infty}\int\frac{d^{3}p}{\beta\left(2\pi\right)^{3}}\left(\gamma_{0}^{E}\right)_{ab} \widetilde{\mathcal{S}}_{ba}\left(\mathbf{p},\omega_{l}^{F}\right)
\end{eqnarray}
is the density of the thermal average of the Noether charge $\widehat{N}$. Since $n\left(\beta, \mu_{e}\right)$ is an intensive thermodynamic parameter, it can be evaluated through the thermodynamic relation

\begin{eqnarray}
n\left(\beta, \mu_{e}\right) & = & \frac{1}{\beta V}\frac{\partial}{\partial\mu_{e}}\left\{ \ln\left[Z\left(\beta,\mu_{e},V\right)\right]\right\},\label{density of N}
\end{eqnarray}
where $Z\left(\beta,\mu_{e},V\right)$ is the complete grand partition function of the generalized quantum electrodynamics. It is an open problem to compute $Z\left(\beta,\mu_{e},V\right)$ exactly but, for our purposes, it suffices to find an approximation for the limit (\ref{limit of Pi}). For instance, in the lowest order of perturbation theory, the only component of $Z\left(\beta,\mu_{e},V\right)$ that depends on the chemical potential is the partition function for free Fermions, whose logarithm reads \cite{LeBe}

\begin{eqnarray}
\ln\left(Z_{f}\right)
 & = & \frac{V}{\pi^{2}}\intop_{0}^{\infty}dpp^{2}\left\{ \beta\sqrt{p^{2}+m_{e}^{2}} +\ln\left[1+e^{-\beta\left(\sqrt{p^{2}+m_{e}^{2}}-\mu_{e}\right)}\right] +\ln\left[1+e^{-\beta\left(\sqrt{p^{2}+m_{e}^{2}}+\mu_{e}\right)}\right]\right\} .\nonumber\\
\end{eqnarray}

In this equation, $m_e$ is the electron mass. Using this in (\ref{density of N}) yields

\begin{align}
n\left(\mu_e,\beta\right)&\,=\frac{1}{\pi^2}\int_0^\infty dp\,p^2\left\{\frac{1}{e^{\beta\left(\sqrt{p^2+m_e^2}-\mu_e\right)}-1} -\left[\frac{1}{e^{\beta\left(\sqrt{p^2+m_e^2}+\mu_e\right)}-1}\right]\right\}.
\end{align}

Unfortunately, there is no known closed form for this integral in terms of elementary functions. However, if we consider the ultrarelativistic regime, that is, the limit in which the energies involved are much higher than the electron rest energy, we find the result from (\ref{limit of Pi}) \cite{Kapusta}

\begin{eqnarray}
\lim_{\mathbf{k}\rightarrow\mathbf{0}}\widetilde{\Pi}_{00}\left(\mathbf{k},0\right)
 & = & \frac{2q_e^2}{\pi^2\beta^2}\left[\mbox{Li}_2\left(-e^{-\beta\mu_e}\right) +\mbox{Li}_{2}\left(-e^{\beta\mu_e}\right)\right]\nonumber\\
 &\simeq & -\frac{q_e^2}{3\beta^2}-\frac{q_e^2\mu_e^2}{\pi^2},\label{ultrarelativistic}
\end{eqnarray}
where $\mbox{Li}_n\left(z\right)$ is the Jonqui\`{e}re's function

\begin{equation}
  \mbox{Li}_n\left(z\right)\equiv \sum_{k=1}^\infty \frac{z^k}{k^n}.
\end{equation}

Instead of the ultrarelativistic regime of (\ref{limit of Pi}) we can consider the other end: its nonrelativistic limit, which is

\begin{eqnarray}
\lim_{\mathbf{k}\rightarrow\mathbf{0}}\widetilde{\Pi}_{00}\left(\mathbf{k},0\right)
 & = & - \frac{\sqrt{2}q_e^2m_e^{\frac{3}{2}}}{\pi^{\frac{3}{2}}\sqrt{\beta}} \left[\mbox{Li}_{\frac{1}{2}}\left(-e^{-\beta\left(m_e-\mu_e\right)}\right) +\mbox{Li}_{\frac{1}{2}}\left(-e^{-\beta\left(m_e+\mu_e\right)}\right)\right]\nonumber\\
 & \simeq & - \frac{\sqrt{2}q_e^2m_e^{\frac{3}{2}}}{\pi^{\frac{3}{2}}\sqrt{\beta}}e^{-\beta m_e}\cosh\left(\beta\mu_e\right).\label{nonrelativistic}
\end{eqnarray}
This last approximation coincides with the result obtained using the Maxwell-Boltzmann statistics.

By performing the Wick rotation (\ref{Wick rotation}) in the complete thermodynamic Green function (\ref{complete Green function}) in order to obtain the retarded propagator, substituting the result for the relevant components of (\ref{potential for Debye}) (\textit{id est}, $\mu=\nu=0$), and using, as previously mentioned, (\ref{point charge}), we find

\begin{eqnarray}
\mathcal{A}_{0}\left(\mathbf{x}\right) & = & Q\int\frac{d^{3}k}{\left(2\pi\right)^{3}} \frac{e^{-i\mathbf{k}\cdot\mathbf{x}}}{\left[\left(1+\frac{\left|\mathbf{k}\right|^{2}}{m_{P}^{2}}\right)\left|\mathbf{k}\right|^{2} -F\left(\mathbf{k},0\right)\right]}.\label{almost electrostatic Debye}
\end{eqnarray}

The long-range behavior of the electrostatic potential in the plasma is governed by the poles of the integrand in the infrared limit of $F$. Taking that into account, we define the Debye mass $m_D$:

\begin{equation}
  m_D\equiv \sqrt{-\lim_{\mathbf{k}\rightarrow\mathbf{0}}F\left(\mathbf{k},0\right)}.\label{Debye mass}
\end{equation}

Due to the identity (\ref{F and Pi}), we see that the computation of $m_D$ is intimately related to the limit $\lim_{\mathbf{k}\rightarrow\mathbf{0}}\widetilde{\Pi}_{00}\left(\mathbf{k},0\right)$, which can be calculated in a number of approximations including, but not limited to, (\ref{ultrarelativistic}) and (\ref{nonrelativistic}).

By exchanging $F\left(\mathbf{k},0\right)$ by its infrared limit $\lim_{\mathbf{k}\rightarrow\mathbf{0}}F\left(\mathbf{k},0\right)$ in (\ref{almost electrostatic Debye}) and defining\footnote{The results presented in this section are valid for $m_D\leq m_P/2$.}

\begin{eqnarray}
m_{D\pm} & \equiv & \sqrt{\frac{m_{P}^{2}\pm m_{P}\sqrt{m_{P}^{2}-4m_{D}^{2}}}{2}}\label{mD plus minus}
\end{eqnarray}
we find the electrostatic potential in the generalized quantum electrodynamics with Debye screening:
\begin{eqnarray}
\mathcal{A}_{0}\left(\mathbf{x}\right)
 & = & \frac{Qm_{P}^{2}}{4\pi\left|\mathbf{x}\right|\left(m_{D+}^{2}-m_{D-}^{2}\right)} \left(e^{-m_{D-}\left|\mathbf{x}\right|}-e^{-m_{D+}\left|\mathbf{x}\right|}\right).\label{Debye potential}
\end{eqnarray}

Comparing these results with those found on the previous section we see that there is a mathematical analogy between the electrostatic potential in the Bopp-Podolsky theory arising from the mass generation \textit{via} Higgs mechanism (\ref{electrostatic Higgs}) and the electrostatic potential of the theory shielded due the interaction with the generalized quantum plasma (\ref{Debye potential}). We also notice that not only the two potentials have the same overall form but also that even the modified masses (\ref{mH plus minus}) and (\ref{mD plus minus}) have the same dependence on the Higgs mass $m_H$ and the Debye mass $m_D$, respectively. So, from the point of view of the electrostatic potential in the Podolsky theory, the Higgs mechanism and the Debye screening are mathematically analogous. For that reason, all the analysis done in the previous section, including all the discussions concerning the results (\ref{limit mH going to zero mHplus}-\ref{approximation    mHminus}), apply to the Debye screening as well. For that, it suffices to make the changes $m_H\rightarrow m_D$ and $m_{H\pm}\rightarrow m_{D\pm}$. Also, all the conclusions we arrive at from now on for the Debye shielding can be translated to Higgs mechanism simply by using the inverse mapping $m_D\rightarrow m_H$ and $m_{D\pm}\rightarrow m_{H\pm}$. For instance, taking into account the approximations (\ref{approximation   mHplus}) and (\ref{approximation    mHminus}), we can recover the Maxwell's result for the Debye screening:\footnote{The limit (\ref{Debye in Maxwell}) holds at least for the lowest order of perturbation theory for $n\left(\beta,\mu_e\right)$.}

\begin{equation}
\lim_{m_P\rightarrow\infty}\mathcal{A}^0\left(\mathbf{x}\right) = \frac{Q\, e^{-m_D\left|\mathbf{x}\right|}}{4\pi\left|\mathbf{x}\right|}.\label{Debye in Maxwell}
\end{equation}

\begin{figure}
  \centering
  \includegraphics[width=10cm]{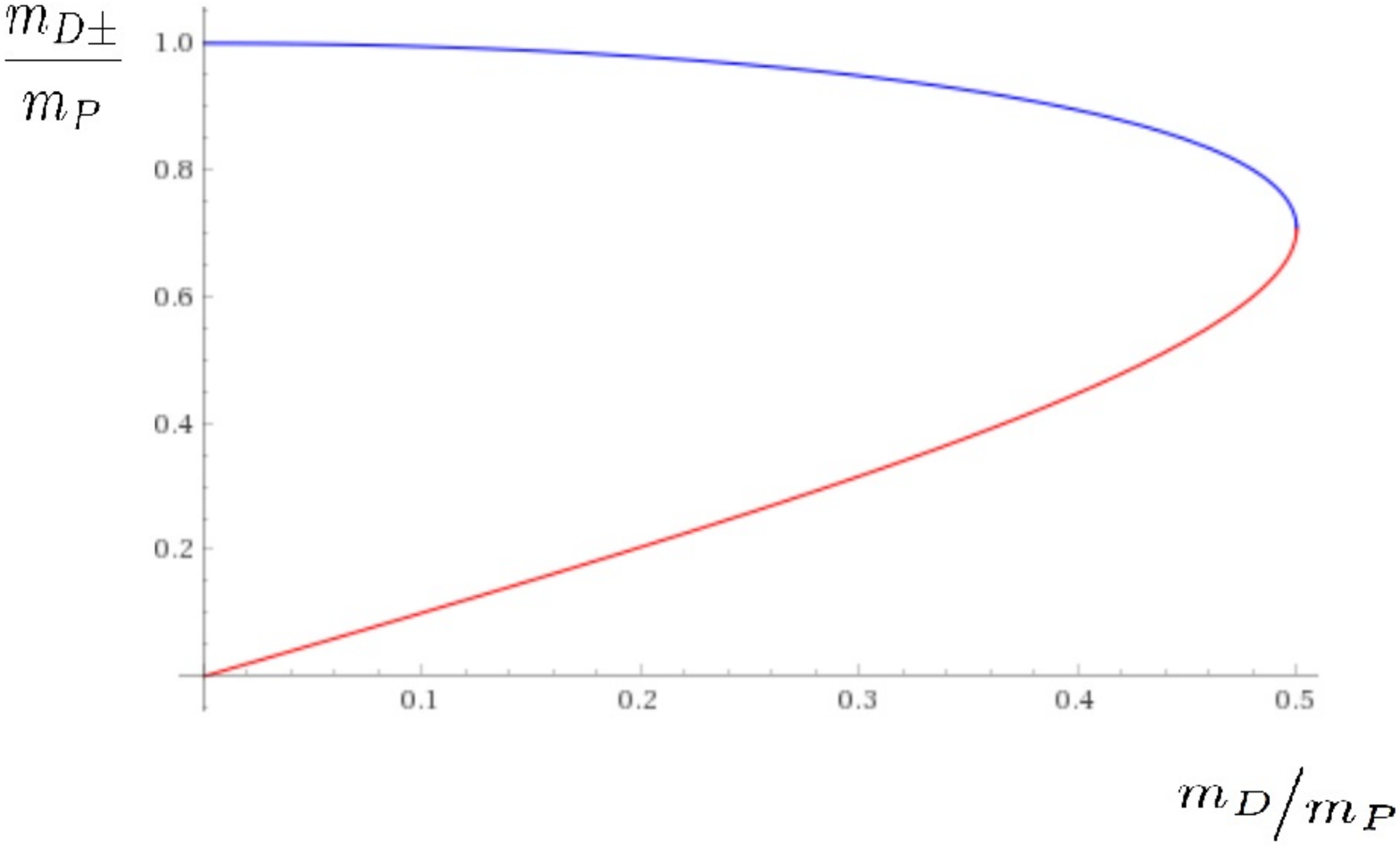}
  \caption{Plot of $m_{D-}/m_P$ (red) and $m_{D+}/m_P$} (blue) as function of $m_D/m_P$\label{Pic the masses}
\end{figure}

Furthermore, by changing the two independent thermodynamic intensive parameters in $n\left(\beta,\mu_e\right)$ we can vary the value of the Debye mass $m_D$. In figure \ref{Pic the masses} we can find the plots of the modified masses $m_{D-}$ and $m_{D+}$ as a function of the Debye mass. In that picture we can visualize the limits (\ref{limit mH going to zero mHplus}) and (\ref{limit mH going to zero mHminus}) and, in the other end, the limit of $m_D\rightarrow \left(m_P/2\right)^-$, which implies $m_{D+}=m_{D-}=m_P/\sqrt{2}$, as discussed in (\ref{equal renormalized}) and (\ref{equal evanescent}).

Notwithstanding, figure \ref{Pic the masses} shows us an important feature of the Debye shielding in the Podolsky theory: by changing the value of the thermodynamic parameters and, as a consequence, changing the value of the original Debye mass $m_D$, we can effectively reduce the value of the mass of the massive sector of the generalized theory from its original value $m_P$ down to $m_P/\sqrt{2}$, which can make it more accessible experimentally. Moreover, even the dependence of the massless sector with the newly acquired mass $m_{D-}$ is not a linear function of the Debye mass $m_D$ as it is in the Maxwell theory. Thus, the previously massless sector, too, can be used as a mean to detect the signature of the elusive Podolsky field.

\begin{figure}
  \centering
  \includegraphics[width=10cm]{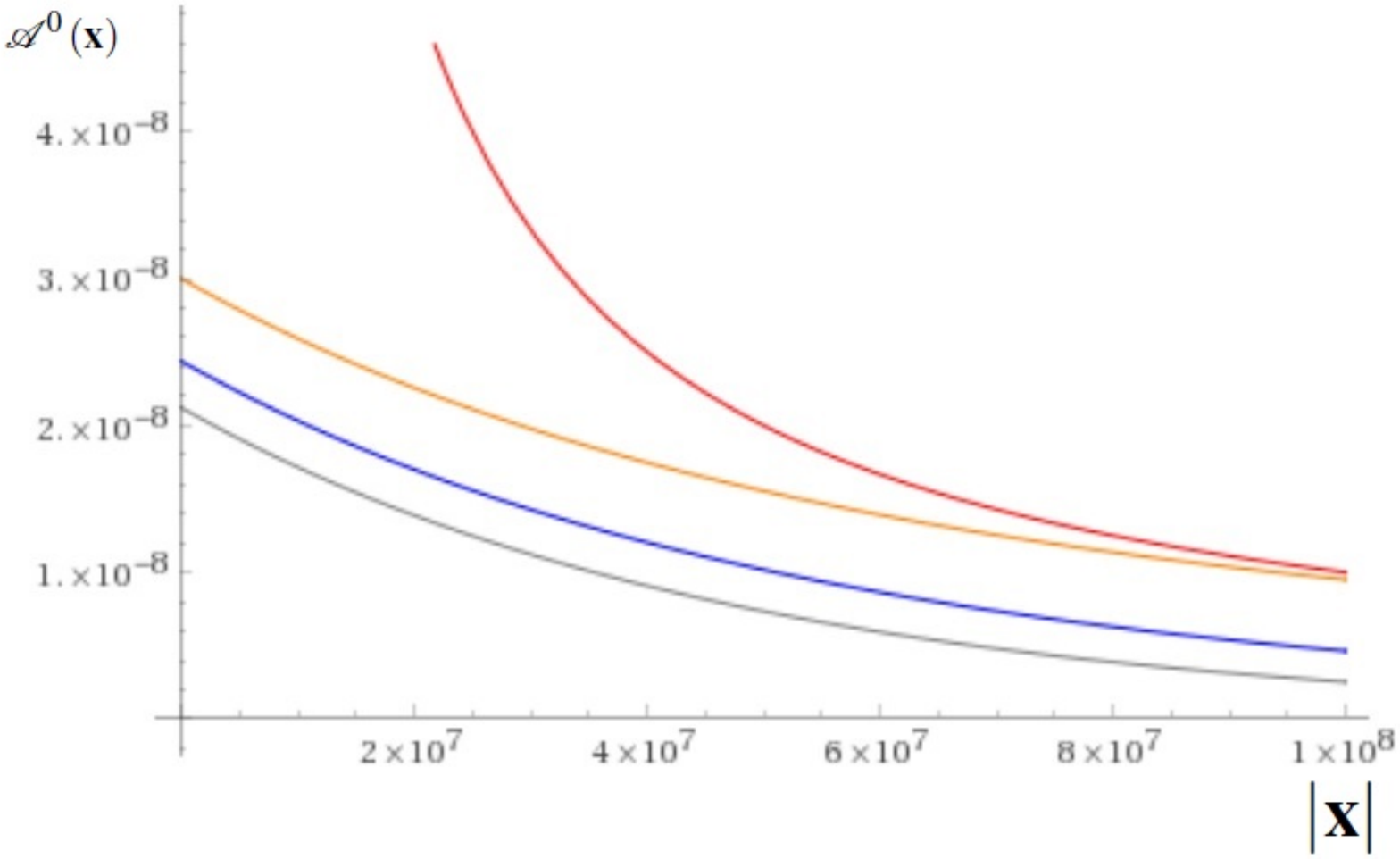}
  \caption{Plot of the electrostatic potential $\mathcal{A}^0\left(x\right)$ as function of the distance $\left|\mathbf{x}\right|$ for $Q=4\pi$ for the Coulomb potential (red), the Podolsky potential (\ref{electrostatic potential in Podolsky}) (orange), and the potential (\ref{Debye potential}) for $m_D=m_P/4$ (blue) and $m_D=m_P/2$ (grey) }\label{Pic with Coulomb}
\end{figure}

\begin{figure}
  \centering
  \includegraphics[width=10cm]{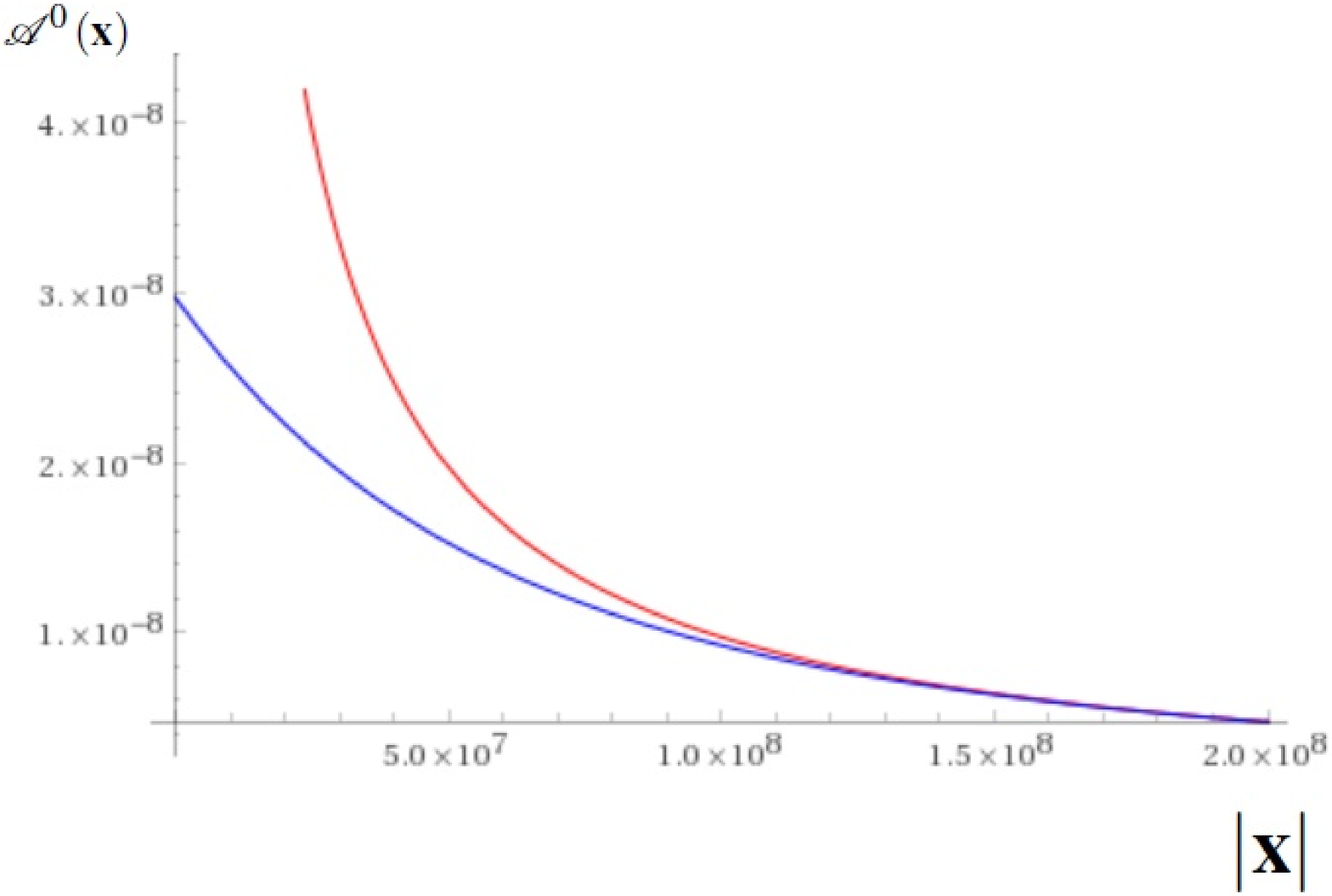}
  \caption{Plot of the screened electrostatic potential for Maxwell (red) and Podolsky (blue) both for $Q=4\pi$ and $m_D=m_P/100$}\label{Pic Debye}
\end{figure}

\begin{figure}
  \centering
  \includegraphics[width=10cm]{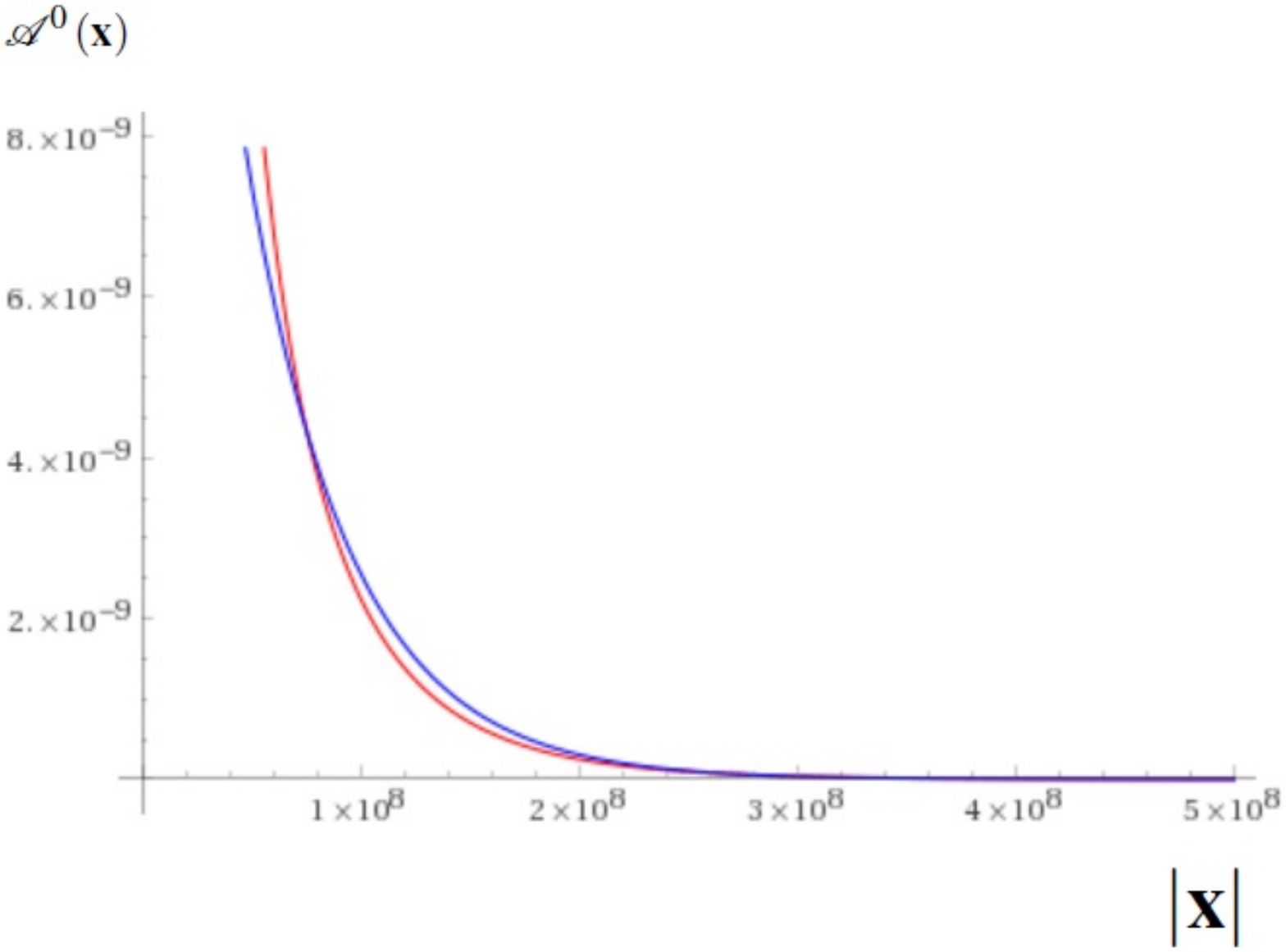}
  \caption{Plot of the screened electrostatic potential for Maxwell (red) and Podolsky (blue) both for $Q=4\pi$ and $m_D=m_P/2$}\label{Pic surprise}
\end{figure}

For the sake of comparison, in figure  \ref{Pic with Coulomb}  we have plotted the Coulomb potential $Q/4\pi\left|\mathbf{x}\right|$, the original Podolsky electrostatic potential (\ref{electrostatic potential in Podolsky}), and the shielded electrostatic potential (\ref{Debye potential}) for $m_D=m_P/4$ and for $m_D=m_P/2$, which happens to be (\ref{equal renormalized}) or (\ref{equal evanescent}).\footnote{In the figures \ref{Pic with Coulomb}, \ref{Pic Debye}, and \ref{Pic surprise} we have used the best minimum fit for the Podolsky mass, which is around $370$ GeV \cite{Daniel}.} We immediately see that (\ref{electrostatic potential in Podolsky}) approaches the Coulomb potential as the distance to the electric charge increases, but the screened Podolsky potential does not (for small distances).\footnote{$\mathcal{A}^0=0$ is a horizontal asymptote for all potentials showed in figure \ref{Pic with Coulomb}. So, \textit{technically}, all those potentials approach Coulomb for large enough distances.} The reason for that is that it does not need to. Instead, the shielded Podolsky electrostatic potential is expected to go to the correspondent screened Maxwell electrostatic potential at long distances. This is, apparently, the case, as can be verified in the example depicted in figure \ref{Pic Debye}, where we used a Debye mass with $1\%$ of the value of the Podolsky mass. That picture also illustrates the finite limit (\ref{finite limit modified}) for the Podolsky electrostatic potential in the quantum plasma. However, this is not entirely accurate. Except for the special case $m_D=m_P/2$ (which shall be dealt with next), we always have $m_{D-}<m_{D+}$. Since $m_{D-}^{-1}$ and $m_{D+}^{-1}$ are, respectively, the typical length of attenuation of the (previously) massless and the massive sectors of the generalized theory, we see that the long range behaviour of the electrostatic potential in the Podolsky theory under Debye screening is dominated by $Qe^{-m_{D-}\left|\mathbf{x}\right|}/4\pi\left|\mathbf{x}\right|$. This is not, however, equal to Debye-screened Coulomb potential $Qe^{-m_{D}\left|\mathbf{x}\right|}/4\pi\left|\mathbf{x}\right|$, unless the approximation (\ref{approximation    mHminus}) holds. In general, though, far away from the source, both the Maxwellian and the Podolskian electrostatic potentials are exponentially attenuated and the farther from the source the harder it is to detect the difference between them.

The situation becomes more interesting, though, when the value of the Debye mass is half of that of the Podolsky mass. In figure \ref{Pic surprise} we have presented the graph of (\ref{equal evanescent}) and that of the Yukawa potential with typical length $2/m_P$, which is the Debye length for $m_D=m_P/2$. The picture shows that, close to the electric source, the absolute value of the Maxwell's results is greater than Podolsky's. This is expected, since the Yukawa potential diverges at the origin, while (\ref{equal renormalized}) does not. What is quite surprising here is the existence of a region of distance where the absolute value of the screened generalized electrostatic field is greater than that of the shielded Coulomb potential. Since both potentials are continuous functions of the distance, this means they intercept at the points satisfying

\begin{equation}
\frac{e^{-\frac{m_P}{2}\left|\mathbf{x}\right|}}{\left|\mathbf{x}\right|} = \frac{m_P\, e^{-\frac{m_P}{\sqrt{2}}\left|\mathbf{x}\right|}}{\sqrt{2}}.
\end{equation}

The solutions for this transcendental equation are

\begin{equation}
\left|\mathbf{x}\right| = -\frac{2\left(1+\sqrt{2}\right)W_n\left(\frac{1}{\sqrt{2}}-1\right)}{m_P},\label{W}
\end{equation}
where $W$ is the Lambert $W$-function, defined as the inverse of $f\left(W\right)=We^W$. The index $n$ in $W$ is an integer that labels the branches of the function. The two interceptions shown in \ref{Pic surprise} are

\begin{align}
\left|\mathbf{x}\right| \simeq &\, \frac{2.2569}{m_P};\\
\left|\mathbf{x}\right| \simeq &\, \frac{8.86008}{m_P},
\end{align}
which are both of the order of $10^{-17}$ m. This is far too small to any current technology's experimental setup to detect. For comparison, this scale is $2$ orders of magnitude smaller than the experimental effective quark-charge radius \cite{HERA}. The kind of inversion of values between Maxwell and Podolsky results showed in \ref{Pic surprise} is unexpected but not unheard of. For instance, the classical Podolsky magnetostatic field is famous to have its value \textit{flipped} in comparison with Maxwell's really close to a infinite, straight wire \cite{Paulo}.

In the next section we shall go even higher in the energy scale.


\section{Beyond the Podolsky Mass}\label{beyond}

In the previous sections we considered the impact the Higgs mechanism has on the Podolsky electrostatic potential as well as the Debye screening in the generalized theory. In both cases, we restrict ourselves to the region $m_H\leq m_P/2$ or, equivalently, $m_D\leq m_P/2$. As a consequence, all the contribution due to the massive sector of the theory acted like corrections (some small, some big) to the usual Maxwellian results. In the present section, on the other hand, we will explore the region of parameters where $m_0> m_P/2$, where $m_0$ stands for either $m_H$ or $m_D$. We will show that in this region the change due to generalized gauge field changes greatly the behaviour of the static potential between electric charges and cannot be thought of as a simple correction. Let us start by considering the Lagrangian density

\begin{eqnarray}
\mathcal{L}_{B} & \equiv & -\frac{1}{4}{F}_{\mu\nu}{F}^{\mu\nu} +\frac{1}{2m_{P}^{2}}\partial^{\mu}{F}_{\mu\nu}\partial_{\xi}{F}^{\xi\nu} +\frac{m_0^{2}}{2}{A}_{\mu}{A}^{\mu}-{J}^{\mu}{A}_{\mu}.\label{last lagrangian}
\end{eqnarray}

In order to explore some phenomenological aspects due to the presence of the nonvanishing mass $m_0$, we start our analysis with the following transition amplitude

\begin{eqnarray}
Z[J]&=&\int DA\exp\left(i\int d^{4}x {\cal L}_{\left(B\right)}\right);\\
{\cal L}_{\left(B\right)}&=&-\frac{1}{4}{F}_{\mu\nu}{F}^{\mu\nu} +\frac{1}{2m_{P}^{2}}\partial^{\mu}{F}_{\mu\nu}\partial_{\xi}{F}^{\xi\nu} -\frac{1}{2}\partial_{\mu}A^{\mu}\left(\frac{\Box}{m_P^{2}}+1\right)\partial_{\nu}A^{\nu} +\frac{m_0^{2}}{2}A_{\mu}A^{\mu}-J_{\mu}A^{\mu}\label{transition amplitude}
\end{eqnarray}
which is the generating functional associated with the quantum version of (\ref{last lagrangian}). We can write the previous transition amplitude in the following way

\begin{eqnarray}
Z[J]&=&\int DA\exp\left[i\int d^{4}x \left(\frac{1}{2}A^{\mu}O_{\mu\nu}A^{\nu}-J_{\mu}A^{\mu}\right)\right];\\
O_{\mu\nu}&\equiv &\eta_{\mu\nu}\left[\left(\frac{\Box}{m_P^{2}}+1\right)\Box+m_0^{2}\right].
\end{eqnarray}
which, after a field translation,

\begin{equation}
A_{\mu}\rightarrow A_{\mu}+\int d^{4}yO^{-1}_{\mu\nu}(x,y)J^{\nu}(y),
\end{equation}
reads

\begin{equation}
Z[J]=\det(O)\exp\left[-\frac{i}{2}\int d^{4}y d^{4}zJ^{\mu}(y)D^B_{\mu\nu}(y-z)J^{\nu}(z)\right].
\end{equation}

Here, we have the propagator

\begin{equation}
\widetilde{D}^B_{\mu\nu}(p)=\frac{m_P^{2}\eta_{\mu\nu}}{(p^{2}-m_P^{2})p^{2}+m_0^{2}m_P^{2}}.
\end{equation}

Now we can investigate how the presence of external sources has influence on the energy of the system. We recall that the transition amplitude (\ref{transition amplitude}) is the vaccum expected value of the time evolution operator $\exp \left\{-i\int_{-\infty}^{+\infty} \widehat{H}\left[J,t\right]dt\right\}$, where $\widehat{H}\left[J,t\right]$ is the Hamiltonian operator associated with the Lagrangian density (\ref{last lagrangian}). So, if we are interested in knowing the change in the energy of the quantum system in the vaccum state due to the presence of the classical sources, we must compute  \cite{Zee, LW}

\begin{equation}
\ln \left(\frac{\left\langle \Omega \left| e^{-i\int_{-\infty}^{+\infty} \widehat{H}\left[J,t\right]dt} \right| \Omega \right\rangle}{\left\langle \omega \left| e^{-i\int_{-\infty}^{+\infty} \widehat{H}\left[0,t\right]dt} \right| \omega \right\rangle}\right) = \ln \left(\frac{Z\left[J\right]}{Z\left[0\right]}\right),\label{log}
\end{equation}
where $\left|\Omega\right\rangle$ and $\left|\omega\right\rangle$ are the vacua states with and without sources, respectively, and $\widehat{H}\left[0,t\right]$ is the Hamiltonian without sources where $Z\left[0\right]$ is its correspondent transition amplitude. Let us consider the following classical source

\begin{equation}
J_{\mu}\left(\mathbf{y}\right)=\eta_{\mu 0}\left[q\delta^{3}\left(\mathbf{y}\right)+Q\delta^{3}\left(\mathbf{y}-\mathbf{x}\right)\right].
\end{equation}

With this, (\ref{log}) becomes $-i\int_{-\infty}^{+\infty}E_T\left(q,Q;\mathbf{x}\right)dt$ where $E_T\left(q,Q;\mathbf{x}\right)=E_{se}\left(q,Q;\mathbf{x}\right)+E_{int}\left(q,Q;\mathbf{x}\right)$ with

\begin{align}
E_{se}\left(q,Q;\mathbf{x}\right)\equiv &\,\frac{q^2m_P^2}{2} \int \frac{d^3p}{\left(2\pi\right)^3}\frac{1}{\left[\left|\mathbf{p}\right|^2\left(\left|\mathbf{p}\right|^2 +m_P^2\right)+m_0^2m_P^2\right]}\nonumber\\
&\,+\frac{Q^2m_P^2}{2} \int \frac{d^3p}{\left(2\pi\right)^3}\frac{1}{\left[\left|\mathbf{p}\right|^2\left(\left|\mathbf{p}\right|^2 +m_P^2\right)+m_0^2m_P^2\right]}
\end{align}
being the self-energy of the sources and

\begin{align}
E_{int}\left(q,Q;\mathbf{x}\right)&\,\equiv qQm_P^2 \int \frac{d^3p}{\left(2\pi\right)^3}\frac{e^{i\mathbf{p}\cdot\mathbf{x}}} {\left[\left|\mathbf{p}\right|^2\left(\left|\mathbf{p}\right|^2 +m_P^2\right)+m_0^2m_P^2\right]}
\end{align}
their interaction (potential) energy. We are interested in this quantity. As it is well-known, the electrostatic potential can be obtained from the potential energy between two charges just by dividing the energy by the electric charge of the test charge. Doing this in $E_{int}$ yields

\begin{equation}
\frac{E_{int}\left(q,Q;\mathbf{x}\right)}{q} = Qm_P^2\int \frac{d^3p}{\left(2\pi\right)^3} \frac{e^{i\mathbf{p}\cdot\mathbf{x}}}{\left[\left|\mathbf{p}\right|^2\left(\left|\mathbf{p}\right|^2 +m_P^2\right)+m_0^2m_P^2\right]}.\label{last potential}
\end{equation}

If $m_0\leq m_P/2$, then we reobtain the results (\ref{electrostatic Higgs}) and (\ref{Debye potential}). However, if $m_0> m_P/2$, the situation becomes way more interesting. For instance, the poles of the integrand of (\ref{last potential}) are $\kappa_b+im_B$, $-\kappa_B+im_B$, and their complex conjugates, with

\begin{align}
m_B &\,\equiv \sqrt{m_0m_P}\cos\left[\frac{1}{2}\arctan\left(\sqrt{\left(\frac{2m_0}{m_P}\right)^2-1}\right)\right]; \label{m b}\\
\kappa_B &\,\equiv \sqrt{m_0m_P}\sin\left[\frac{1}{2}\arctan\left(\sqrt{\left(\frac{2m_0}{m_P}\right)^2-1}\right)\right]. \label{kappa b}
\end{align}

There are a few things to notice about (\ref{m b}) and (\ref{kappa b}). Firstly, since we are in the regime $m_0>m_P/2$, we have $\sqrt{\left(\frac{2m_0}{m_P}\right)^2-1}>0$. As a consequence

\begin{equation}
0<\frac{1}{2}\arctan\left(\sqrt{\left(\frac{2m_0}{m_P}\right)^2-1}\right)<\frac{\pi}{4}.
\end{equation}

Therefore, both $\kappa_B$ and $m_B$ are positive. In this case, the electrostatic potential (\ref{last potential}) is

\begin{equation}
A\left(\mathbf{x}\right) = \frac{Q\, m_P^2}{8\pi} \frac{e^{-m_B\left|\mathbf{x}\right|}}{m_B} \frac{\sin\left(\kappa_B\left|\mathbf{x}\right|\right)}{\kappa_B\left|\mathbf{x}\right|}.\label{drastic}
\end{equation}

The first thing we notice is that for $m_0$ greater than half of the Podolsky mass value there is a drastic change in the phenomenology of the generalized electrostatic potential. It goes from a Yukawa-like potential to an oscillatory, although attenuated, one. As a matter of fact, (\ref{drastic}) can be viewed as a sine function with period $2\pi/\kappa_B$ enveloped by the Yukawa potential $Q\,e^{-m_B\left|\mathbf{x}\right|}/4\pi\left|\mathbf{x}\right|$ and rescaled by the multiplying constant $m_P^2/2m_B\kappa_B$. Consequently, the potential periodically vanishes at the points $\left|\mathbf{x}\right|=n\pi/\kappa_B$, $n\in\mathbb{N}^+$. Just to give us a glimpse of the distance between these zeroes, let us suppose $m_0\sim m_P$. Using the best lower bound for the Podolsky mass \cite{Daniel}, it would furnish $m_B\sim m_P$ and $\kappa_B\sim m_P$ and we would find the zeroes for the potential equidistanced from each other at the order $10^{-17}$ m, which is well-beyond anything experimental currently available. The second thing to notice is the limit

\begin{equation}
\lim_{m_0\rightarrow \frac{m_P}{2}^+}A^0\left(\mathbf{x}\right)=\frac{\sqrt{2}Q\, m_P}{8\pi}e^{-\frac{m_P}{\sqrt{2}}\left|\mathbf{x}\right|},
\end{equation}
which happens to coincide with the result for equal masses (\ref{equal continuous}). Therefore, the screened Podolsky electrostatic potential is continuous everywhere as a function of the underlying mass $m_0$. The third thing to notice is that even in this regime of energy, the potential still has a finite limit at the origin:

\begin{equation}
\lim_{\left|\mathbf{x}\right|\rightarrow 0^+}A^0\left(\mathbf{x}\right)= \frac{Q}{8\pi}\frac{m_P^2}{m_B}.
\end{equation}

But now, a question arises: which is the regime of energy where this drastic change in the potential takes place? In order to answer that question it is convenient to go back to the Debye mass specifically. In the ultrarelativistic limit (\ref{ultrarelativistic}) in the situation of vanishing chemical potential, the Debye mass, due to equations (\ref{F and Pi}) and (\ref{Debye mass}), reads

\begin{equation}
m_D=\frac{\sqrt{4\pi\alpha}T}{\sqrt{3}},
\end{equation}
where we have written the electron electric charge in terms of the fine structure constant $\alpha$ and $T=\beta^{-1}$ is the temperature. By equating $m_D$ to $m_P/2$ and using the best lower experimental value available for the Boop-Podolsky parameter, arising from the uncertainties of the gyromagnetic ratio or electron-positron scattering measures \cite{Daniel}, we find a temperature of the order $T_B\sim 10^{23}$ K. Just to give us an idea of the scale we are dealing with, the classical radius of the electron is approximately ($r_e\sim 10^{-15}m$), to which we can associate a temperature through $m_ec^{2}=k_BT_e$ of the order $T_e\sim 10^{9}$ K. Although $T_B$ is a formidable temperature and beyond anything ever produced by humans, this temperature may not be beyond the reach of probing after all. As a matter of fact, this is a mere order of magnitude lower than the temperature associated with one specific observation by the Fly's Eye air shower detector \cite{OMG}. That observation consisted of an ultra-high-energy cosmic ray event whose energy is around $3.2\times 10^{20}$ eV, which translates as $3.7\times 10^{24}$ K.

We used the relation (\ref{log}) to compute the vacuum energy change due to classical sources in the quantum Bopp-Podolsky field instead of simply computing the electrostatic potential of the theory to elucidate the fact that at this scale of energy there is no hope of a classical description producing anything close to accurate predictions. However, even the formulation presented here's aim is nothing more than to give us a pedagogical insight of what kind of change to expect in the gauge field behaviour alone at so high temperatures. A more realistic description, though, should not neglect the contributions from the other fundamental interactions of the Standard Model of Particles.


\section{Conclusions}\label{Conclusions}

In the present work we studied two mechanisms for mass generation in the Podolsky theory: the Higgs mechanism and the Debye screening. Although the Higgs mechanism can be studied in a more general context, our main focus rested in the electrostatic potential of the theory. The reasons are twofold: first of all, even the electrostatic regime contains all the pertinent features of the mass generation through the symmetry breaking, which makes its generalization to include other  field configurations (like time-dependent ones) possible. And, secondly, it allows us to compare our results with those of Debye shielding, which takes place with static electric fields. From this comparison we found the most important result of this paper: in what concerns the electrostatic potentials, the two mechanisms for mass generation are mathematically analogous in their results. Perhaps the importance of this result is overlooked when one thinks of Maxwell: both mechanisms generate (effective) masses for the Maxwell field. Notwithstanding, what studying a second-order derivative theory teaches us is not only that there is generation of masses for the gauge field making the electrostatic potentials (\ref{electrostatic Higgs}) and (\ref{Debye potential}) look alike, but also that, independently of the underlying mechanism, the masses change in exactly the same way in both effects, as can be seen in equations (\ref{mH plus minus}) and (\ref{mD plus minus}). This mathematical analogy is kept secret when one is studying Maxwell electrodynamics, but it is revealed in its fullness when studying the generalized theory.

Furthermore, we analised limits and the behaviours of the static potentials in the parameter region $m_0\leq m_P/2$. We found that by carefully changing the independent, intensive parameters, we can change the value of the Debye mass (or, equivalently, by somehow changing the coupling constant between the Podolsky field and the Higgs field, we can change the Higgs mass) which allows us to change the value of changed masses (\ref{mD plus minus}). This is helpful when trying to probe the existence of the Podolsky parameter since by raising the value of $m_D$ it simultaneously lowers the effective value of Podolsky mass $m_{D+}$ as well as changes the form of the function of newly acquired mass $m_{D-}$.

Lastly, we saw that in the previously mentioned regime, all that the contribution from the massive sector of the Bopp-Podolsky theory does is ``correcting" the Maxwell's electrostatic potential in some sense. A substantial changing in the qualitative behaviour of the potential takes place at energies above the threshold $m_P/2$, though. From that point on, the potential behaves like a standing sine wave enveloped by a rescaled Yukawa potential. The transition between these two regions is continuous and its short-range limit is still finite. Moreover, we estimated the order of magnitude for the temperature for this change in the potential to happen. This temperature, although very high indeed, is not extremely beyond the reach of physical observation, being only one order of magnitude below that associated with the most energetic cosmic ray event detected to date. We hope that future observations of events like that shed some light in the behavior of the electromagnetic field.

Although we have studied how the infra-red mass generating mechanisms (Higgs or Debye shielding) affect the Podolsky electrodynamics, it is possible to explore other types of mechanisms \cite{Stueck, BGAP}. This matters will be analysed and require elaborations.

\section*{Acknowledgement}
CAB thanks his family for support in this time of crisis and the hospitality of UEPG's 105 Group, G. B. de Gracia thanks CAPES PhD grant (CP), A. A. N. thanks National Post-Doctoral Program grant (PNPD) for support and B. M. P. thanks CNPq for partial support.

\end{document}